\documentclass[11pt]{article}

\usepackage[margin=1in]{geometry}

\usepackage{graphicx}
\graphicspath{{../../graphics/},{../manuscript_1_17_2017/}}
\usepackage{amsmath,amssymb,amsfonts,amsthm}
\usepackage[round,sort&compress]{natbib}

\usepackage[lined,ruled,linesnumbered]{algorithm2e}
\SetKw{KwSet}{Set}
\SetKw{KwIterate}{Iterate}
\SetKwInOut{Input}{input}\SetKwInOut{Output}{output}

\usepackage{hyperref}
\hypersetup{backref,colorlinks=true,citecolor=blue,linkcolor=blue,urlcolor=blue}
\usepackage{superPCA}

\usepackage{mathrsfs}

\newcommand{\indicator}{\boldsymbol{1}}
\newcommand{\A}{\mathbf{M}}
\newcommand{\F}{\mathbf{F}}
\newcommand{\I}{\mathbf{I}}
\renewcommand{\S}{\mathbf{S}}

\usepackage{grffile}

\usepackage{booktabs}

\title{Predicting phenotypes from
  microarrays using amplified, initially marginal, eigenvector regression}
\author{Lei Ding\\Department of Statistics\\Indiana University, Bloomington, IN\\\email{dinglei@indiana.edu}\and 
Daniel J. McDonald\\Department of Statistics\\Indiana University, Bloomington, IN\\\email{dajmcdon@indiana.edu}}

\begin{document}

\maketitle

\begin{abstract}
\noindent\textbf{Motivation:}
  The discovery of relationships between gene expression
  measurements and phenotypic responses is hampered by
  both computational and statistical impediments. Conventional statistical methods
  are less than ideal because they either fail to select relevant
  genes, predict poorly, ignore the unknown interaction structure
  between genes, or are computationally intractable. Thus, the
  creation of new methods which can handle many expression 
  measurements on relatively small numbers of patients while
  also uncovering gene-gene relationships and predicting well is desirable.\\
 \noindent \textbf{Results:} We develop a new technique for using the marginal
  relationship between gene expression measurements and patient
  survival outcomes to identify a small subset of genes which appear
  highly relevant for predicting survival, produce a low-dimensional embedding based on this
  small subset, and amplify this embedding with information from
  the remaining genes. We motivate our methodology by using
  gene expression measurements to predict survival time for patients with diffuse large B-cell
  lymphoma, illustrate the behavior of our
  methodology on carefully constructed synthetic examples, and test it
  on a number of other gene expression datasets. Our technique is
  computationally tractable, 
  generally outperforms other methods, is extensible to other
  phenotypes, and also
  identifies different genes (relative to existing methods) for
  possible future study.\\
 \noindent \textbf{Key words:} regression; principal components;
  matrix sketching; preconditioning\\
  \noindent\textbf{Availability:} All of the code and data are available at
  \url{https://github.com/dajmcdon/aimer/}.
\end{abstract}

\section{Introduction}

\label{sec:introduction}


A typical scenario in genomics is to obtain expression
measurements for thousands of genes from microarrays or RNA-Seq which may be
relevant for predicting a particular phenotype. Such studies have been useful in relating specific genetic
variations to a wide variety of outcomes such as disease specific
indicators~\citep{LesageBrice2009,barrett2008genome,burton2007genome,SladekRocheleau2007};
drug or vaccine
response~\citep{SaitoIkeda2016,KennedyOvsyannikova2012}; and individual
traits like motion 
sickness~\citep{HromatkaTung2015} or age at
menarche~\citep{ElksPerry2010,PerryDay2014}.
In these scenarios, researchers are interested in the accurate
prediction of the phenotype and the identification of a handful of
relevant genes with a reasonable computational expense. With these goals in mind, supervised linear regression
techniques such as ridge regression~\citep{hoerl1970ridge}, the lasso
~\citep{Tibshirani1996}, the Dantzig selector~\citep{CandesTao2007},
or other penalized methods are often employed. 

However, because phenotypes tend to be the result of groups of genes,
which perhaps together describe more complicated biomechanical processes,
rather than individual polymorphisms, recent approaches have tried to
account for this group structure. Techniques such as the group lasso~\citep{YuanLin2005} can predict the
response with sparse groupings of coefficients as long as the
groups are partially understood ahead of time. In contrast,
unsupervised methods such as principal components
analysis~\citep{hotelling1957relations,Jolliffe2002,pearson1901principal} are often used directly on the
genes when no phenotype is being
examined~\citep{AlterBrown2000,SladekRocheleau2007,WallRechtsteiner2003}.
Finally, modern approaches developed specifically for the genomics
context such as supervised gene shaving~\citep{HastieTibshirani2000},
tree harvesting~\citep{HastieTibshirani2001}, and supervised principal
components~\citep{bair2004semi,bair2006prediction} have sought to
combine the presence of a response with the structure estimation properties of eigendecompositions
from unsupervised techniques to obtain
the best of both. It is this last set of techniques that most closely resemble the
approach we present here. We give a more detailed discussion of
supervised principal components next, before motivating our
method with an example.

\textbf{Notation:} We will use bolded letters $\mathbf{M}$ to indicate
matrices, capital letters to denote column vectors, such that $M_j$ is
the $j^{th}$ column of the matrix $\mathbf{M}$, and lower case letters
$m_i$ to denote row vectors (a single subscript) or scalars ($m_{ij}$ being the
$i,j$ element of $\mathbf{M}$). We will use 
the notation $\mathbf{M}_A$ to mean the columns of $\mathbf{M}$
whose indices are in the set $A$ and $[k]=\{1,\ldots,k\}$. 
Finally, for a matrix $\mathbf{M}$, we
write the singular value decomposition (SVD) of
$\mathbf{M}=\mathbf{U(M)}\boldsymbol{\Lambda}(\mathbf{M})
\mathbf{V(M)}^\top$ and define $\A^\dagger$ to be the Moore-Penrose
inverse of $\A$. In the case only of the design matrix
$\mathbf{X}$ discussed below, we will use the more compact
decomposition $\mathbf{X}   =   \mathbf{U}   \boldsymbol{\Lambda}
\mathbf{V}^\top$. 

\subsection{Supervised eigenstructure techniques}
\label{sec:superv-eigenstr-tech}

The first technique for extending unsupervised principal components
analysis to the case where a response is available is principal
components regression (PCR,
\citealp{hotelling1957relations,kendall1965course}). Instead of
regressing the response on all the available
covariates as in ordinary least squares (OLS), PCR first performs an eigendecomposition of the empirical
covariance matrix and then regresses the response on the subset of principal
components corresponding to the largest variances. Defining
$Y\in\R^n$ to be the centered response vector, and
$\mathbf{X}$ to be the $n \times p$ centered design matrix, write the
(reduced) SVD of $\mathbf{X}$ as  
$\mathbf{X}   =   \mathbf{U}   \boldsymbol{\Lambda}    \mathbf{V}^\top.$
For some integer $d\leq p$, the principal components
regression estimator is 
given as the solution to
\begin{align*}
 \boldsymbol{\hat{\Gamma}}_{PCR} = \argmin_\Gamma \norm{Y-\mathbf{U}_{[d]}
\boldsymbol{\Lambda}_{[d]}\boldsymbol{\Gamma}}_2^2,
\end{align*}
which has the closed form representation
\begin{align*}
        \boldsymbol{\hat{\Gamma}}_{PCR} 
        & = ((\mathbf{U}_{[d]}  \boldsymbol{\Lambda}_{[d]})^\top
          \mathbf{U}_{[d]}   \boldsymbol{\Lambda}_{[d]})^{-1} (
          \mathbf{U}_{[d]}  \boldsymbol{\Lambda}_{[d]})^\top Y 
         = \boldsymbol{\Lambda}^{-1}_{[d]}   \mathbf{U}^{T}_{[d]}   Y.
\end{align*}
Since this solution is in the space spanned by the principal components, it is
easy to rotate the estimate back onto the span of $\mathbf{X}$:
$\boldsymbol{\hat{\beta}}_{PCR}:=\mathbf{V}_{[d]}
\boldsymbol{\hat{\Gamma}}_{PCR} =
\mathbf{V}_{[d]}\boldsymbol{\Lambda}^{-1}_{[d]}   \mathbf{U}^{T}_{[d]}
Y$. Then any elements of $\boldsymbol{\hat{\beta}}_{PCR}$
which are identically zero imply the irrelevance of those  genes for predicting the
phenotype while the columns of $\mathbf{V}_{[d]}^\top$ can be interpreted
as indicating groupings of individual genes.

Principal components regression performs well under certain conditions
when we believe that there are 
natural groupings of covariates (linear combinations) which are useful
for predicting the response. However, \cite{lu2002sparse} 
and \cite{johnstone2004sparse} show that the empirical singular vectors
$\mathbf{U}_{[d]}$ are poor estimates of the associated population
quantity (the left singular vectors of the expected value of
$\mathbf{X}$) unless $p/n \to 0 $ as
$n \to \infty$. In particular, when $p\gg n$, as is common in genomics 
where the number of gene expression measurements is much larger
than the number of patients, PCR will suffer.

To avoid this flaw in PCR, various approaches have been
proposed. \citet{HastieTibshirani2000} proposed a method called ``gene
shaving'' that is applicable to both supervised (given a
phenotype) and unsupervised (only gene expressions) settings. In the
supervised setting, it works by computing the first principal
component and ranking the genes using a combined measure that
balances the principal component scores and the marginal relationship
with the response. Those genes with lowest combined scores are removed
and the process is repeated until only one gene remains, resulting in
a nested sequence of clusters containing fewer
and fewer genes. Then one chooses a cluster along this sequence,
orthogonalizes the data with respect to the genes in that cluster, and
repeats the entire process again, iterating until the desired number
of clusters has been recovered. This procedure is somewhat
computationally expensive as well as requiring both the 
cluster sizes and the number of clusters to be chosen.

An alternative with somewhat similar behavior is
supervised principal components (SPC,  
\citealp{bair2004semi,bair2006prediction}).  SPC avoids the
high-dimensional regression problem 
by first selecting a much smaller subset of useful genes which have
high marginal
correlation with the phenotype
(in contrast to gene shaving, which uses the marginal correlation and
the covariance between genes). By screening out most of the hopefully
irrelevant genes, we can return to the scenario where $p<n$. In
follow-up work, \citet{paul2008preconditioning} show that, 
if a small marginal correlation with the response implies
irrelevance for prediction, then SPC will find any truly relevant
genes and predict the phenotype accurately. They also suggest
using lasso or forward stepwise selection after SPC to further reduce the number of genes. 
However, if some genes have
small marginal relationship with the response but large conditional
relationship, they will be erroneously ignored by SPC. It is this last
property that our method attempts to correct. We now illustrate that
the screening step of SPC is likely to remove important genes in
typical applications before discussing how our procedure
avoids suffering the same fate.

\subsection{A motivating example}
\label{sec:motivating-example}
\begin{figure}[t]
\centering
\includegraphics[width=.5\linewidth]{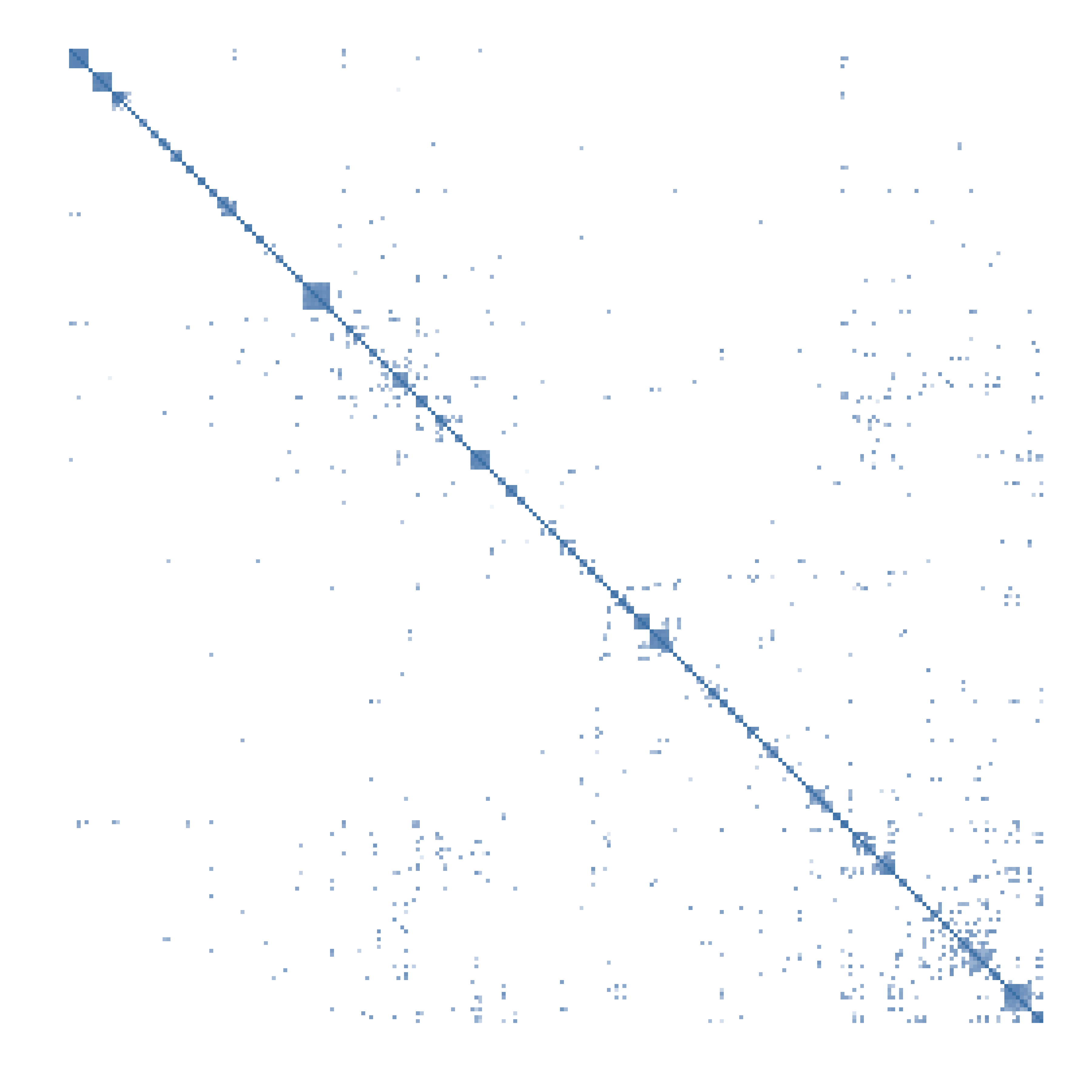}
\caption{A sparse estimate of the inverse covariance of gene
  expression measurements for the first 250 genes from the DLBCL
  dataset. The estimate has 97.5\% of the off-diagonal elements equal to
 0. Darker colors represent inverse covariances of larger magnitude.}
\label{fig:sparse-icov}
\end{figure}

To motivate our methodology in relation to previous approaches, we
examine a dataset consisting of 240 patients with diffuse large B-cell
lymphoma~\citep[DLBCL,][]{rosenwald2002use} in some detail. Each patient is
measured on 7399 genes, and her survival time is recorded. 
\begin{table}[t]
  \centering
  \resizebox{\textwidth}{!}{
\begin{tabular}{@{}lcrrrrrrrrrr@{}}
\toprule
  sparsity of $\boldsymbol{\Sigma}_{xx}^{-1}$&\phantom{ab}& 1.0000 & 0.9999 & 0.9998 & 0.9995 & 0.9991 & 0.9984 &
                                                                 0.9975
  & 0.9963 & 0.9946 & 0.9922 \\  
  \midrule
  \% non-zero $\beta$'s & &0.0162 & 0.0216 & 0.0287 & 0.0418 & 0.0618 & 0.0843 &
                                                                      0.1193 & 0.1803 & 0.2645 & 0.3699 \\  
  False Negative Rate & &0.0000 & 0.2500 & 0.4340 & 0.6117 & 0.7374 &
                                                                     0.8077
                                                                                                   & 0.8641 & 0.9100 & 0.9387 & 0.9562 \\   
  \bottomrule
\end{tabular}
}
\caption{This table shows properties of the coefficients of the linear
  model corresponding to 10 different estimates of the inverse
  covariance matrix, from complete sparsity on the left (a diagonal
  matrix) to still more than 99\% sparsity on the right. The second
  row is the number of non-zero population regression coefficients
  corresponding to each inverse covariance matrix. The bottom row
  shows  the percentage of non-zero regression coefficients which are
  incorrectly ignored under 
  the assumption on the relationship between marginal correlations and 
  regression coefficients.}
\label{tab:assumption}
\end{table}
Previous
approaches rely on the assumption that a small marginal correlation between
the response variable, in this case patient survival time, and the
vector of expression measurements for a particular gene is sufficient
for guaranteeing the irrelevance of that particular gene for
prediction. To make this assumption mathematically precise, suppose 
$
y = x^\top \beta + \epsilon,
$
where $y$ is the response, $x$ is a vector of gene expression
measurements, and $\epsilon$ is a mean-zero error. Then, the assumption can be stated mathematically as
$\textrm{Cov}(x_{j}, y)=0 \Rightarrow \beta_j=0$. While reasonable
under some conditions, this assumption is perhaps too strong for many
gene expression datasets. Very often, individual gene expressions are
only predictive of phenotype in the presence of other genes. We can
rewrite this assumption using the population 
covariance matrix between genes, $\textrm{Cov}(x,x) = \boldsymbol{\Sigma}_{xx}$,
and the vector-valued covariance
between gene expressions and phenotype,
$\textrm{Cov}(x,y)=\Sigma_{xy}$. Then, using the population equation
for $\beta$ allows us to rewrite the assumption as
\begin{equation}
(\Sigma_{xy})_j=0 \Rightarrow
\beta_j=(\boldsymbol{\Sigma}_{xx}^{-1}\Sigma_{xy})_j=0.
\label{eq:assumption}
\end{equation}
In words, we are assuming that the dot product of the $j^{th}$ row of
the inverse covariance matrix with the covariance between $x$ and $y$
is zero whenever the $j^{th}$ element of $\Sigma_{xy}$ is zero. 

To examine whether this assumption holds, we can 
estimate both $\boldsymbol{\Sigma}^{-1}_{xx}$ and $\Sigma_{xy}$ using
the DLBCL data and imagine that these estimates are the population
quantities for illustration. 
To estimate $\Sigma_{xy}$, we use the standard covariance estimate,
but set all but the largest 120 values equal to zero, corresponding to a sparse
solution. For the case of $\boldsymbol{\Sigma}^{-1}_{xx}$, estimating large inverse
covariance matrices accurately is impossible when $p\gg 
n$ unless we assume some additional structure. If most
of the entries are 0 (a necessary condition for \eqref{eq:assumption} to hold),
methods like the graphical lasso~\citep[glasso,][]{FriedmanHastie2008}
or graph estimation~\citep{MeinshausenBuhlmann2006} have been shown to work 
well. We use the graph estimation technique for all 7399 genes in the
dataset at ten different sparsity levels ranging from 100\% to
99.2\%. For visualization purposes, \autoref{fig:sparse-icov} shows
the first 250 genes for one estimate of the inverse covariance that is
97.5\% sparse.  

To assess the validity of \eqref{eq:assumption}, \autoref{tab:assumption} shows the
sparsity of the full inverse covariance matrix, the
percentage of non-zero regression coefficients, and the percentage of
non-zero regression coefficients which are incorrectly ignored by the
assumption (the false negative rate). In all cases, $\Sigma_{xy}$ is
about 98\% sparse. Even with an extremely sparse
inverse covariance matrix, the false negative rate is at least 25\%
meaning that 25\% of possibly relevant genes are ignored by the analysis. If the
sparsity of $\mathbf{\Sigma}^{-1}_{xx}$ is allowed to increase only slightly, the false negative rate
increases to over 95\%.

\subsection{Our contribution}
\label{sec:ourContribution}
For a similar computational budget, our method outperforms existing
approaches by taking advantage of all the data. Our method does not
require that the set of non-zero regression coefficients be a subset
of the non-zero marginal correlations.

Suppose that $\A \in \mathbb{R}^{p\times p}$ is a symmetric,
nonnegative definite matrix; that is, for all vectors $a \in \mathbb{R}^{p}$,
$a^{\top} \A a \geq 0$ and $\A^{\top} = \A$. To approximate the matrix
$\A$, we fix an integer $\ell \ll p$ and form a sketching matrix $\S \in
\R^{p \times \ell}$.  Then, we report the following approximation: $\A
\approx (\A\S)(\S^{\top}\A\S)^{\dagger}(\A\S)^{\top}$.  The details
behind the formation of the matrix $\S$ control the type of
approximation.

In the simplest case, which we employ here, we take 
  $\S = \pi\tau, $
where $\pi \in \R^{p\times p}$ is a permutation of the identity matrix
and $\tau = [\I_\ell, \mathbf{0}]^{\top} \in \R^{p \times \ell}$ is a
truncation matrix.  While many alternative sketching matrices, mostly based on random
projections, have been proposed, this method is the only one necessary
to develop our results.
Without loss of generality, divide the matrix $\A$ into blocks
\begin{equation*}
  \A = 
    \begin{bmatrix}
      \A_{11} & \A_{21}^{\top} \\
      \A_{21} & \A_{22}   \\
    \end{bmatrix}
\end{equation*}
so that we can (implicitly) construct the matrix $\F(\A) \in \R^{p \times \ell}$ as
\begin{equation*}
  \F(\A) := \A\S = \left[ \begin{array}{c} \A_{11} \\\A_{21} \\ \end{array} \right].
\end{equation*}

Because
\begin{align*}
  \A
  &\approx (\A\S)(\S^{\top}\A\S)^{\dagger}(\A\S)^{\top}
  = \F(\A) (\S^{\top}\A\S)^{\dagger}\F(\A)^{\top},
\end{align*}
we can approximate the eigendecomposition of $\A$ using the SVD of
$\F(\A)$.  If we decompose $\F
   = \mathbf{U}(\F) \boldsymbol{\Lambda} (\F) \mathbf{V}(\F)^{\top},
$
where we have suppressed the dependence of $\F$ on $\A$ when $\F$ is an argument
for clarity,  then the resulting approximation to the
eigenvectors of $\A$ is 
$
  \mathbf{V}(\A) \approx \F \mathbf{V}(\F) \boldsymbol{\Lambda}(\F)^{\dagger} = \mathbf{U}(\F).
$
Likewise, the approximate eigenvalues of $\A$ are given the
singular values $\boldsymbol{\Lambda}(\F)$.  

\citet{HomrighausenMcDonald2016} show that this approximation is more
accurate than the one based on $\A_{11}$ for performing a
principal components analysis. As previous techniques for principal
components regression (like SPC) are based on $\A_{11}$ rather than $\F$, it is
possible that by using $\F$, we will have better results. As we will
see, this intuition turns out to be true under some conditions
which were suggested in~\autoref{sec:motivating-example}. In
particular, for essentially the same computational budget, our
procedure outperforms previous procedures if some genes have small
marginal correlations with the phenotype but are, nonetheless, important
for predicting the phenotype conditional on the presence of other
genes. Furthermore, even if the assumption in \eqref{eq:assumption} is
true, our procedure is not much 
worse than existing approaches. 

In \autoref{sec:computing}, we discuss exactly how to implement our
methodology. We examine the behavior of our procedure in 
\autoref{sec:empiricalResults}.  In \autoref{sec:notation-model}, we
state an explicit model for 
the data-generating mechanism in order to be clear about the
conditions under which our procedure works
well. \autoref{sec:experiments} uses a number of carefully
constructed simulations to show when our technique works well, and
when it doesn't. In \autoref{sec:empiricalExample}, we examine our procedure on four 
genetics datasets, including the one discussed above. We find that our
methods slightly outperform existing techniques on three of them,
suggesting that the motivation is sound. Finally, in
\autoref{sec:discussion}, we give conclusions and discuss some avenues
for future work.

\section{Methods and computations}
\label{sec:computing}



We now give the details of our methodology. For clarity, we assume that the design
matrix $\mathbf{X}$ and the response $Y$ are already
centered. Let $T$ be a $p$-dimensional vector denoting
standardized regression coefficient estimates, i.e. for any $j \in
\{1, 2, \ldots, p\}$, $t_j$ is the coefficient estimate of
standardized univariate regression between response $Y$ and
covariate $X_j$. We use standardized regression so that the
coefficient estimates are comparable across disparate covariates. Note that
$t_j$ is also the marginal correlation between the response
$Y$ and covariate $X_j$. 

For some threshold $t_*$, we separate $\mathbf{X}$ into two
matrices $\mathbf{X}_A$ and $\mathbf{X}_{A^c}$, where $A = \{j:\left| t_j
\right| > t_*\}$. We assume $|A|=\ell$. The hope is that
$\mathbf{X}_A$ contains many of the genes that are most predictive of
the phenotype under study. Ideally, high marginal correlations will
suggest relevant predictors to be emphasized in the decomposition, but
unlike other methods, we will also use those genes in the set $A^c$. We now
focus on $\mathbf{X}_{new} = [ \mathbf{X}_{A}, \ 
\mathbf{X}_{A^c}]$ and note that it has the same range as
$\mathbf{X}$. Therefore, we will use the approximation technique discussed in
\autoref{sec:ourContribution} to try to estimate the
eigendecomposition of $\mathbf{\Sigma}_{xx}$ using sample
quantities.  Because $\mathbf{X}_{new}^\top  \mathbf{X}_{new}$ is
symmetric and positive definite, write
\begin{equation*}
  \F   =   \mathbf{X}_{new}^\top   \mathbf{X}_{A} =
        \begin{pmatrix}
                \mathbf{X}^\top_A    \mathbf{X}_A\\
                \mathbf{X}^\top_{A^c}    \mathbf{X}_A
        \end{pmatrix},
\end{equation*}
and decompose $\F = \mathbf{U}(
\F )  \boldsymbol{\Lambda} ( \F )   \mathbf{V}(
\F  )$. For some integer  $d\in\{1,\ldots,\ell\}$, we define
\begin{align*}
  \mathbf{\hat{V}}_{[d]}
  &=  \mathbf{U}_{[d]}(  \F  ),\\
  \mathbf{\hat{\Lambda}}_{[d]}
  &=  \boldsymbol{\Lambda}_{[d]}(  \F  )^{1/2}, \quad\quad\textrm{and}\\
  \mathbf{\hat{U}}_{[d]}  &=  \mathbf{X}_{new}
\mathbf{\hat V}_{[d]}  \boldsymbol{\hat \Lambda}^{-1}_{[d]}.
\end{align*}

Now we have estimates for the principal components $\mathbf{\hat{U}}_{[d]}
\boldsymbol{\hat{\Lambda}}_{[d]}$. Therefore, just as with principal
components regression, we can regress $Y$ on the estimated principal
components to produce estimated coefficients in principal component space:
\begin{align*}
 \boldsymbol{\hat{\Gamma}}_{AIMER} = \argmin_\Gamma
  \norm{Y-\mathbf{\hat U}_{[d]}
  \boldsymbol{\hat \Lambda}_{[d]}\boldsymbol{\Gamma}}_2^2
        & 
         = \boldsymbol{\hat \Lambda}^{-1}_{[d]}   \mathbf{\hat U}^{T}_{[d]}   Y.
\end{align*}
Then the coefficient estimates for linear regression in the space
spanned by $\mathbf{X}_{new}$ are given by 
\begin{equation}
        \boldsymbol{\hat{\beta}}_{AIMER} 
        =   \mathbf{\hat{V}}_{[d]}   \boldsymbol{\hat{\Gamma}}_{AIMER} 
        =   \mathbf{\hat{V}}_{[d]}   \boldsymbol{\hat{\Lambda}}^{-1}_{[d]}
        \mathbf{\hat{U}}^{T}_{[d]}    Y  .
        \label{eq:unreg-estimate}
\end{equation}
Because our methodology uses marginal regression to select a small
number of hopefully relevant predictors before ``amplifying'' their
eigenstructure information with the $\F$ matrix, we refer to our technique as
``Amplified, Initially Marginal, Eigenvector Regression'' (AIMER).

Unlike previous approaches, the solution given by
\eqref{eq:unreg-estimate} is not sparse: with probability 1,
$(\boldsymbol{\hat{\beta}}_{AIMER})_j \neq 0,\ \forall j$.
However, most of the coefficients will be small. We therefore
threshold the estimates to produce our final estimator:
\begin{equation}
  \label{eq:final-estimator}
  \boldsymbol{\hat{\beta}}_{AIMER}(b) :=
  \boldsymbol{\hat{\beta}}_{AIMER} \indicator_{(b,\infty)}(|
  \boldsymbol{\hat{\beta}}_{AIMER}|),
\end{equation}
where $b \geq 0$, and $\indicator_A(w)$ is the indicator function, which returns the
value one for every element of $w\in A$ and zero otherwise. We
summarize this procedure in \autoref{alg:estimation}. As with 
SPC, the computational burden of our method is dominated by the
SVD. We use an SVD of $\F$ while SPC uses the SVD of 
$\mathbf{X}_A$. However, since the SVD is cubic in the smaller
dimension, in both cases the computation is $O(|A|^3)$. Thus, to
leading order, both methods require the same amount of computation.

\begin{algorithm}[t]
  \KwIn{centered design matrix $\mathbf{X}$, centered response
    $Y$, thresholds $t_*,b_*\geq 0$, integer $d$}
   Compute marginal correlation $t_j$ between $X_j$ and $Y$ for all $j$\;
  
  \KwSet $A=\{ j : \left| t_j \right| > t_*\}$\; 

  \KwSet $\mathbf{X}_{new} = [ \mathbf{X}_{A}, \  \mathbf{X}_{A^{c}}]$\;

  Define $\F  =  \mathbf{X}_{new}^\top  \mathbf{X}_{A}$\;

  Decompose $\F = \mathbf{U(\F)\Lambda(\F)V(\F)}^\top$\;

  \KwSet $\mathbf{\hat{V}}_{[d]}  =  \mathbf{U}_{[d]}(\F)$\; 

  \KwSet $\boldsymbol{\hat{\Lambda}}_{[d]}  =  \boldsymbol{\Lambda}_{[d]}(\F)^{1/2}$\;

  \KwSet $\mathbf{\hat{U}}_{[d]}  =  \mathbf{X}_{new}  \mathbf{\hat V}_{[d]}
  \boldsymbol{\hat \Lambda}^{-1}_{[d]}$\; 

  Calculate $\boldsymbol{\hat{\beta}}  =  \mathbf{\hat{V}}_{[d]}
  \boldsymbol{\hat{\Lambda}}^{-1}_{[d]}    \mathbf{\hat{U}}^{T}_{[d]}
  \mathbf{Y}$ \;

  \KwSet  $\boldsymbol{\hat{\beta}}(b_*) :=
  \boldsymbol{\hat{\beta}} \indicator_{(b_*,\infty)}(|
  \boldsymbol{\hat{\beta}}|)$\;

  \KwOut{coefficient estimates $\boldsymbol{\hat{\beta}}(b_*)$}    
  \caption{Amplified, Initially Marginal, Eigenvector
  Regression (AIMER) }\label{alg:estimation}
\end{algorithm}

To make predictions given a new observation $x_*$,
we simply center it using the mean of the original data, reorder its entries to
conform to $\mathbf{X}_{new}$, multiply by the coefficient vector in
\eqref{eq:final-estimator}, and add the mean of the original response
vector. 






\section{Experimental analysis}
\label{sec:empiricalResults}

To examine the performance of our method, we set up a number of
carefully constructed simulations under various conditions. We first
discuss the generic data model we assume, a latent factor
model, which is amenable to analysis via SPC or AIMER.


\subsection{Data model}
\label{sec:notation-model}

\begin{figure}
\centering 
\includegraphics[width=\textwidth]{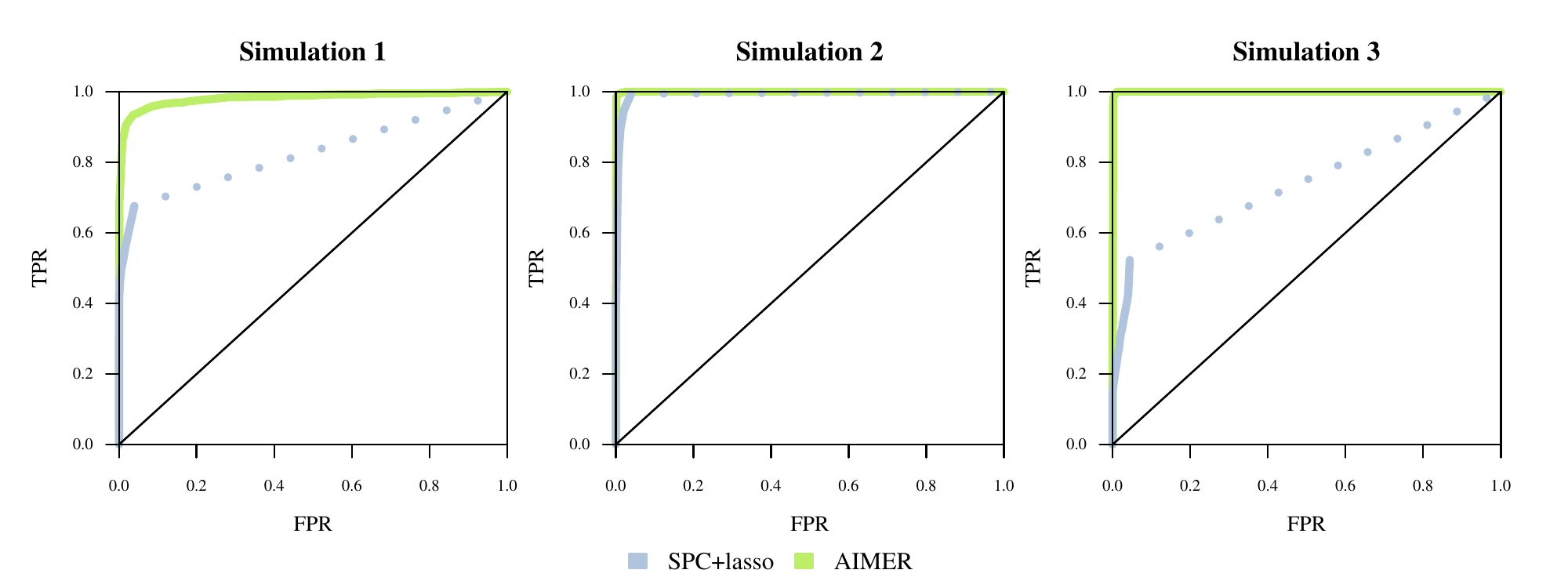} 
\caption{Receiver operating characteristic (ROC) Curve for Simulations 1--3. The $x$-axis is the
  false positive rate while the $y$-axis is the true positive
  rate. The curves present averages across 100 replications. SPC is limited to only 50 selected genes, and so its false
  positive rate is bounded. The dashed line indicates its best case
  theoretical performance were it allowed to continue to select
  further genes.}\label{fig:sim123comparisonROC}
\end{figure}

\begin{figure*}
\centering 
\includegraphics[width=\textwidth]{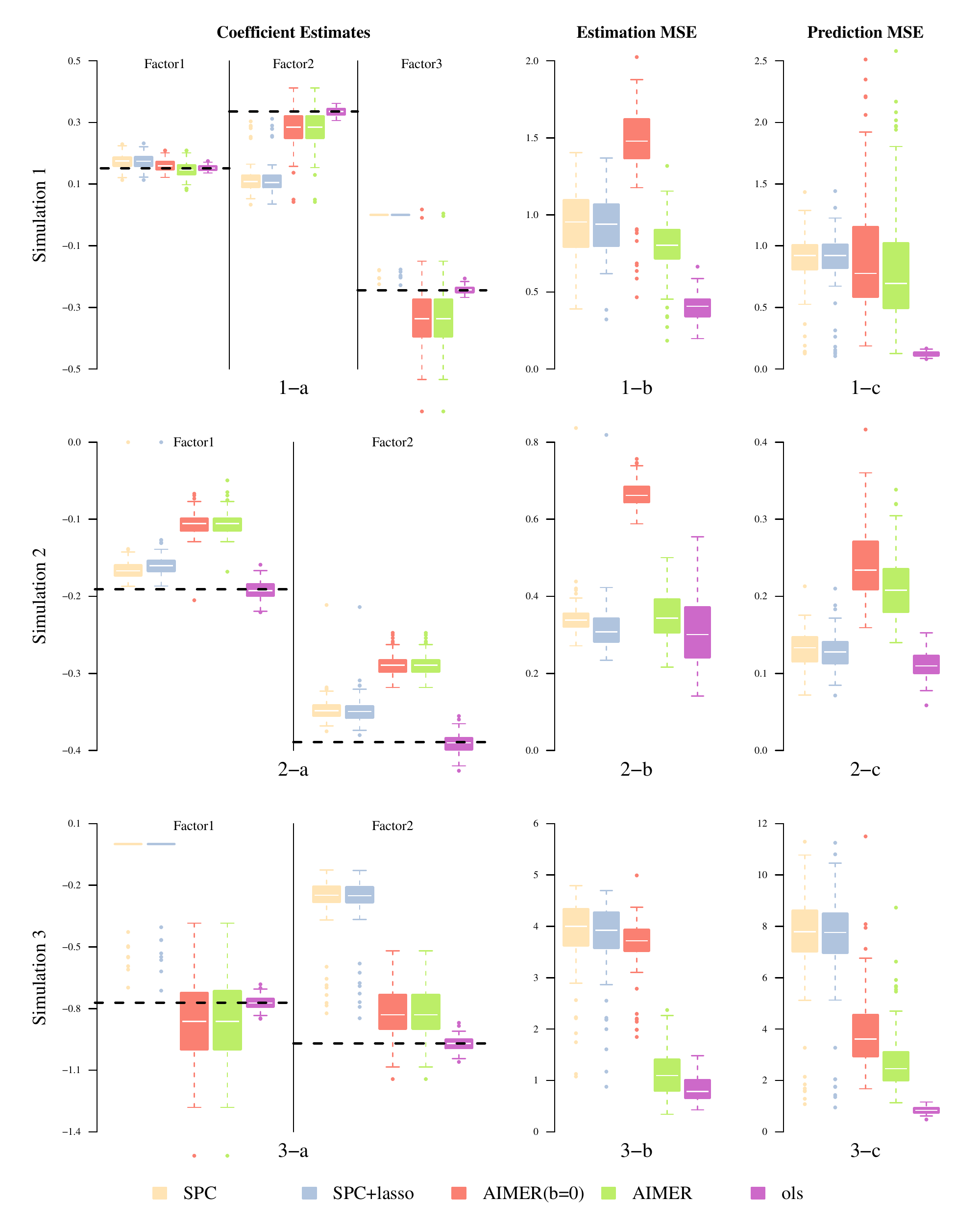} 
\caption{Estimation and prediction performance of SPC and AIMER in the
first three simulations. The left panel shows the estimates of the
regression coefficients, the middle panel shows the mean squared error
(MSE) of estimation
for all 1000 genes, and the right panel shows prediction MSE on the
held-out data. The boxes indicate variability across 100
replications. The dashed black horizontal lines indicate the true
values of $\beta$.}
\label{fig:sim123comparison}
\end{figure*}

\begin{figure*}
\centering 
\includegraphics[width=\textwidth]{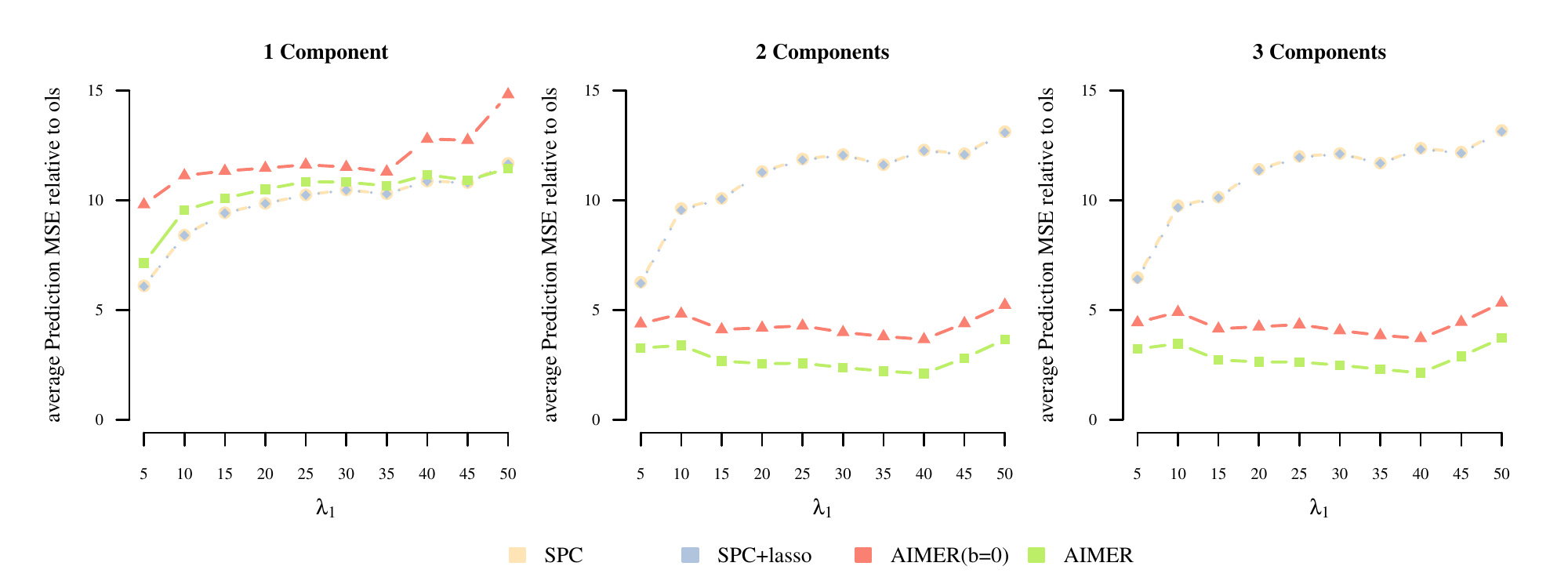} 
\caption{Prediction MSE averaged across 100 replications for each
  method for different numbers of components (Simulation 4). We also allow
  $\lambda_1$ to vary between 5 and 50.}
\label{fig:sim4}
\end{figure*}

\begin{figure*}
\centering 
\includegraphics[width=\textwidth]{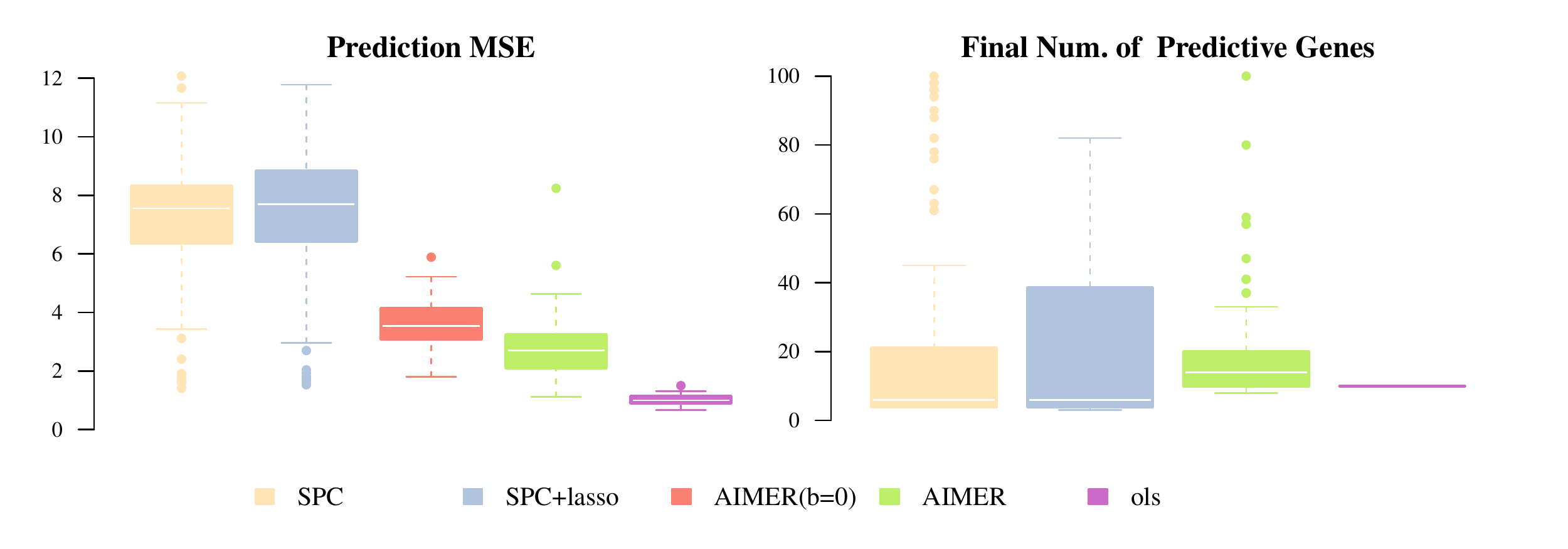} 
\caption{Performance of each method when we allow $t_*$ to be chosen
  by cross validation rather than fixed to choose 50 genes (Simulation
  5).}
\label{fig:sim5}
\end{figure*}

\begin{table}
\centering
\begin{tabular}{@{}lclll@{}}
\toprule
Simulation  &\phantom{ab} &  1  &  2   &  3   \\
\midrule
True \#                 & & 15                  & 10    & 10     \\
SPC               & & 50     \tiny{(0)}               & 50      \tiny{(0)}               & 50       \tiny{(0)}\\
SPC+lasso    & & 31      \tiny{(9.011)}     & 39      \tiny{(3.636)}      & 46       \tiny{(2.665)}\\
AIMER($b=0$)          &        & 1000  \tiny{(0)}              & 1000  \tiny{(0)}               & 1000   \tiny{(0)} \\
AIMER       & & 39      \tiny{(9.225)}     & 21      \tiny{(12.750)}    & 16       \tiny{(7.558)} \\
\bottomrule
\end{tabular}
\caption{\label{tab:2}Average final number of predictive genes in
  Simulations 1, 2, and 3. The standard deviation is shown in parentheses.}
\end{table}

Consider the multivariate Gaussian linear regression model 

\begin{equation}
y = x^\top \boldsymbol{\beta}+\sigma_1\epsilon\label{eq:linear-model}
\end{equation}
with $y$ the
response, ${x}\in\R^p $ a column vector of gene expression measurements,
$\boldsymbol{\beta} = (\beta_1,  \cdots,  \beta_p )^\top$ the
coefficients,  $\epsilon$ a random Gaussian distributed error with
zero mean and variance 1, and $\sigma_1>0$. We further assume that $
x \sim N_p 
(\mathbf{0}, \boldsymbol{\Sigma}_{xx} )$ has a Gaussian distribution with
mean vector $\mathbf{0}$ and covariance matrix $\boldsymbol{\Sigma}_{xx}$. 
%
We will assume that $\boldsymbol{\beta}$ is sparse, in that most of
its elements are exactly 0 indicating no linear relationship between
the associated gene and the response. 
Finally, the design matrix $\mathbf{X}$ and the response vector $Y$
include $n$ independent observations of $x$ and $y$ respectively.

\paragraph{Model for $\mathbf{X}$.} As $\boldsymbol{\Sigma}_{xx}$ is symmetric and positive
(semi-) definite, we can decompose it as
\begin{align*}
\boldsymbol{\Sigma}_{xx}   
  &=   \mathbf{V}(\boldsymbol{\Sigma}_{xx})
    \mathbf{L}(\boldsymbol{\Sigma}_{xx})
    \mathbf{V}^{T}(\boldsymbol{\Sigma}_{xx})\\
    &= 
    \begin{pmatrix}
        V_1 & \cdots & V_p
    \end{pmatrix}
    \begin{pmatrix}
        l_1      & & 0      \\
         & \ddots & \\
        0      & & l_p
    \end{pmatrix}
    \begin{pmatrix}
        V^\top_1  \\
        \vdots \\
        V^\top_p  
    \end{pmatrix},
\end{align*}
where $V_1, \cdots, V_p$ are orthonormal
eigenvectors on $\mathbb{R}^p$ and $l_1 \ge \cdots \ge l_p \ge 0$ are
eigenvalues. We assume that there is some $1\leq G \leq p$ such
that the eigenvalues can be seperated into two
groups, one of which includes relatively large eigenvalues and the other
relatively small eigenvalues, that is, $l_k=\lambda_k+\sigma^2_0$ for
$1\leq k \leq G$ and $l_k=\sigma^2_0$ for $k>G$
%
%
where $\lambda_1 \ge \cdots \ge \lambda_G > 0$, and $\sigma^2_0 > 0.$ 

Then, because $\mathbf{X}$ is multivariate Gaussian, we
can write $\mathbf{X}$ as  
\begin{align*}
    \mathbf{X}  
     &=   
     \mathbf{U}_{G}    \mathbf{\Lambda}_{G}     \mathbf{V}^{T}_{G}   +
       \sigma_0   \mathbf{E}   \notag\\ 
    &=
    \begin{pmatrix}
        U_1 & \cdots & U_G
    \end{pmatrix}
    \begin{pmatrix}
        \sqrt{\lambda_1}      &  & 0      \\
        & \ddots & \\
        0      & & \sqrt{\lambda_G}
    \end{pmatrix}
    \begin{pmatrix}
      V^\top_1  \\
      \vdots \\
      V^\top_G    
    \end{pmatrix}
    + 
    \sigma_0    \mathbf{E} 
\end{align*}
where latent factors $U_1, \ldots, U_G$ are
independent and identically distributed (i.i.d.) $N_{n}(\mathbf{0},
\mathbf{I})$ vectors, and the noise matrix 
$\mathbf{E}$ is $n \times p$ with i.i.d.\ $N(0, 1)$ entries
independent of $U_1, \ldots, U_G$. 

\paragraph{Model for $Y$.} We assume that $Y$ is
a linear function of the first $K\leq G$ latent factors in
$\mathbf{U}_{G}$ plus additive Gaussian noise: 
  $Y = \mathbf{U}_{K}   \boldsymbol{\Theta} +
  \sigma_{1}   Z,$
where $\boldsymbol{\Theta}$ is the
coefficient vector, $\sigma_1 > 0$ is a constant, and $Z$ is
distributed $N_{n}(\mathbf{0}, \mathbf{I})$,
independent of $\mathbf{X}$. Note that the expectation of $Y$
is zero and that this is a specific form of \eqref{eq:linear-model}. 

\paragraph{Implication of the model.} Under this model for
$\mathbf{X}$ and $Y$,
the population marginal covariance between 
each gene $X_j$ and the response $Y$ can be written as
\begin{align}
\label{eq:covariace_X_Y}
  \Sigma_{xy} 
        &= 
        \begin{pmatrix}
                \textrm{Cov}(X_1, Y) \\
                \vdots \\
                \textrm{Cov}(X_p, Y)   
        \end{pmatrix}
        = 
        \mathbf{V}_K     \mathbf{\Lambda}_K    \mathbf{\Theta}. 
\end{align}
Therefore, the population ordinary least squares coefficients of
regressing $Y$ on $\mathbf{X}$ ($\boldsymbol{\beta}$
in \eqref{eq:linear-model}) can be written as
\begin{align}\label{eq:population_regression_coefficient}
        \boldsymbol{\beta} 
        &= 
        \boldsymbol{\Sigma}_{xx}^{-1}    \Sigma_{xy} 
        = 
        \mathbf{V}_K    \mathbf{L}^{-1}_K    \mathbf{\Lambda}_{K}    \boldsymbol{\Theta}  
\end{align}  
We will define the set $\mathscr{B} := \{ j: (\Sigma_{xy})_j \ne 0 \}$
and the set $\mathscr{A} := \{ j: \beta_{j} \ne 0 \}$. We note that
for $K=1$, it is always the case that
$\mathscr{A} = \mathscr{B}$. By manipulating
the parameters in $\boldsymbol{\Theta}$, $\mathbf{L}$, and $\mathbf{\Lambda}$, we
can create a number of scenarios for testing AIMER against alternative
methods.

\subsection{Experiments}
\label{sec:experiments}

We present results under five different experiments. For
each of the simulations which follow, we generate datasets 
with $n=200$ and $p=1000$. We
use half ($n=100$) to estimate the model and test our predictions on
the other half. We repeat this process 100 times for each combination
of parameters. Throughout, we use $\sigma_0=\sqrt{.1}\approx .3$ and
$\sigma_1=.1$. The matrix $\mathbf{U}$ is generated with i.i.d.\
standard Gaussian entries, while the matrix $\mathbf{V}$ is
constructed by hand to have the correct number of orthogonal
components. 

The first experiment is designed to be favorable to AIMER. The second is designed to
be favorable to SPC. The third examines the extent to which the
assumption that $\mathcal{A}=\mathcal{B}$ is beneficial to SPC over
AIMER. The fourth examines the impact of using incorrect numbers of
components, while the fifth uses cross validation on all the tuning
parameters. 

\paragraph{Simulation 1: Favorable conditions for AIMER.}
\label{sec:simulation1}

In this simulation, we create data which is amenable to AIMER at the
expense of the conditions for SPC, that is we use $\mathscr{B} \subset \mathscr{A}$. 
We set parameters in the data model as $K=G=3$ and choose $\lambda_{1}=10,$
$\lambda_{2}=5,$ and $\lambda_{3}=1$. In order to achieve $\mathscr{B} \subset
\mathscr{A}$, we set $\theta_{1}=\theta_2=1$ and solve 
\eqref{eq:covariace_X_Y} for $\theta_{3}$ so that some corresponding
elements of $\Sigma_{xy}$ will be zero.
We make the first 15 elements of $\boldsymbol{\beta}$ non-zero, 5
corresponding to each of the three principal components. Thus,
the first 10 genes have
non-zero population marginal correlation and the remaining 990 have
zero marginal correlation. In this scenario, SPC should find the
first 10 important genes, but AIMER will find the remaining
5 important genes as well.

In order to focus on the relationship between performance and the
condition $\mathscr{B}\subset\mathscr{A}$, we examine the methods for a fixed
computational budget and choose $t_*$ to select the same $50$ most
predictive genes. We examine SPC, SPC with lasso, AIMER($b=0$), and 
AIMER. We use the first 3 principal components for regression in all the methods. 
For SPC with lasso and AIMER, we choose the remaining tuning parameters
via 10-fold cross-validation. We also give results for OLS on the first
15 genes. This is the oracle estimator, the best one could hope to 
do with foreknowledge of the predictive genes.

\autoref{fig:sim123comparisonROC} shows the classification performance
using a receiver operating characteristic (ROC) curve
for SPC with lasso and AIMER in the left panel (the remaining panels
are for the next two simulations). Examining the figure, it is easy to
see that SPC+lasso identifies the first 10 genes easily, 
but AIMER is able to capture all 
15 predictive genes at a low cost of false positive identifications. A
more detailed analysis is given in the first row of
\autoref{fig:sim123comparison}. Panel 1--a shows the ability of each
method to estimate the $\boldsymbol{\beta}$ coefficients of three different factors. 
Coefficient estimates for the 5 genes in factor 1
by AIMER are slightly more accurate, and no more variable, than SPC+lasso.
Furthermore, AIMER is better at estimating those $\boldsymbol{\beta}$'s 
associated with factor 2, 
and much better at those associated with factor 3 (these are
assumed zero in SPC).
Panel 1--b examines the mean square error (MSE) of estimation as the
average squared difference between the true coefficients and their
estimates for all 1000
genes. The overall estimation accuracy of AIMER($b=0$)
is worse
because of the inclusion of so many useless genes (it estimates all 1000), 
however, by thresholding with AIMER, accuracy is improved and
exceeds that of SPC with and without lasso. 
In panel 1--c, we show the MSE for
prediction, the average squared difference between predicted values
and the actual observations, for a test set. This MSE is smaller for AIMER than
for SPC much of the time, 
but the variance across simulations is large.

\paragraph{Simulation 2: Favorable conditions for SPC.}
\label{sec:simulation2}

This simulation compares the performance of SPC and AIMER under
conditions which are more favorable to SPC. In particular, we choose
parameters such that $\mathscr{A}=\mathscr{B}$. While AIMER is likely
to perform worse because it will tend to include irrelevant genes, it
is not too much worse.
Most of the parameters are the same as in Simulation 1,
except that $K=G=2$, $\lambda_{1}=10$,
$\lambda_{2}=1$,   $\theta_{1}=\theta_{2}=1$,
and we use the first two principal components to do regression. 
Therefore, 10 out of 1000 genes are truly
predictive of the response, and all 10 have non-zero marginal
correlation with the response (the rest have $\Sigma_{xy}=0$).
Looking again at \autoref{fig:sim123comparisonROC},  both SPC+lasso and 
AIMER can identify all 10 predictive genes at a small price of false positives.
Examining \autoref{fig:sim123comparison}, we see that the estimation
accuracy of SPC/SPC+lasso is better than that of AIMER
as expected, and the MSE of prediction for AIMER is about twice that
of SPC/SPC+lasso. The estimation MSE (panel 2--b) of AIMER is
comparable to that of SPC.

\paragraph{Simulation 3: Slight perturbations.}
\label{sec:simulation3}

In this simulation, we adjust only $\theta_{2}=3$, rather than 1 as in simulation 2, thereby maintaining
the condition that $\mathscr{A} = \mathscr{B}$.
However, in this case AIMER works much better than SPC/SPC+lasso. 
Figures \ref{fig:sim123comparisonROC} and \ref{fig:sim123comparison} show
that AIMER can easily identify all the predictive genes, has more
precise coefficient estimates, and has much smaller MSE for prediction. 
The reason is that, even though $\mathscr{A} = \mathscr{B}$, 
the marginal correlations for some predictive genes are very small.
Therefore those genes are more difficult for SPC to identify, but AIMER can compensate.

For one further comparison, \autoref{tab:2} shows the average
(standard deviation in parentheses) number of predictive genes 
selected in each of the first three
simulations.  AIMER selects the smallest number of
coefficients in most cases.

\paragraph{Simulation 4: Choosing the number of components.}
\label{sec:simulation4}

\begin{table*}
  \centering
  \resizebox{\textwidth}{!}{
\begin{tabular}{@{\extracolsep{9pt}}l r r r r r r r r r r r r@{}}
\toprule
& \multicolumn{3}{c}{DLBCL}  & \multicolumn{3}{c}{Breast cancer}   & \multicolumn{3}{c}{Lung cancer}  & \multicolumn{3}{c}{AML}\\
\cline{2-4} \cline{5-7} \cline{8-10} \cline{11-13}
Methods  &  MSE  &  \# genes   &  $d$   &  MSE  &  \# genes   &  $d$  &  MSE  &  \# genes   &  $d$  &  MSE  &  \# genes   &  $d$\\
\midrule
lasso            & 0.6805   & 20   &   & \textbf{0.6285}   & 9   &   & 0.8159   & 22   &   & 1.9564   & 6   &   \\
ridge              & \textbf{0.6485}   & 7399   &   & 0.6407   & 4751   &   &
                                                                   \textbf{0.7713}   & 7129   &   & \textbf{1.9234}   & 6283  &   \\
\midrule
SPC                & 0.6828   & 41   & 3   & 0.6066   & 16   & 2   & \textbf{0.8344}   & 19   & 3   & 2.4214   & 24   & 2  \\
SPC+lasso  & 0.6780   & 31   & 3   & 0.6029   & 14   & 2   & 0.8436   & 9   & 4   & 2.3980   & 22   & 2  \\
AIMER($b=0$)                   & 1.1896   & 7399   & 2   & 2.6531   & 4751   & 1   & 0.9444   & 7129   & 1   & 12.4014   & 6283   & 1  \\
AIMER         & \textbf{0.6518}   & 28   & 4  & \textbf{0.6004}   & 31   & 3   & 1.0203   & 13   & 1   &  \textbf{1.8746}   & 36   & 4  \\
\bottomrule
\end{tabular}
}
\caption{\label{tab:1} The MSE on the test set, the number of selected genes, and the
number of principal components used ($d$ if relevant), each averaged
across the 10 random training-testing splits. Bolded values indicate the
best predictive performance for each type of method (with and
without structure learning) for each data set.}
\end{table*}

In the previous simulations, we used the correct number of principal
components, though such a choice is unlikely to be possible given real
data. In this simulation, we examine the impact choosing the number of
components has on estimation accuracy.
We use similar parameter settings as Simulation 1 except with $K=G=2$
rather than 3 (we maintain the condition that
$\mathcal{B}\subset\mathcal{A}$). We then use all the methods 
with 1, 2, and 3 components. We also adjust the values of
$\lambda_{1}$ in a range from 5 to 50. 
As we can see in \autoref{fig:sim4}, 
using two components reduces MSE for AIMER($b=0$) and AIMER 
across all values of
$\lambda_1$ relative to using only one component, while
using more than two components has little impact. With only one
component, SPC performs better than AIMER, likely due to smaller
variance for a similar bias, but using two or three components leads
to large gains for AIMER. In practice, 
it is worthwhile to try several numbers of components and use
cross-validation to decide which works best.

\paragraph{Simulation 5: The screening threshold.}
\label{sec:simulation5}

In previous simulations, we choose $t_*$ so that variable screening by
the marginal correlation would always select
exactly 50 genes. Thus, we could compare methods based on their
ability to use the same amount of information. In reality, it may be better to choose the
threshold $t_*$ using cross validation. 
In this simulation, we use the same conditions as in the previous
simulation with $\lambda_1=10$. It is still not appropriate to have
more genes than patients,  
so we allow the number of selected genes to be anything less than the
number of patients (100). 
We further use 10-fold cross-validation to choose the best threshold.

As shown in \autoref{fig:sim5}, allowing $t_*$ to be chosen rather
than fixed leads to improved results for AIMER relative to
SPC/SPC+lasso. The prediction MSE decreases and fewer
genes are selected.

\section{Performance on real data}
\label{sec:empiricalExample}

We now illustrate our methods on 4 empirical datasets in genomics that
record the censored 
survival time and gene expression measurements from DNA microarrays of
patients with 4 different types of cancer. The first dataset comes
from \citet{rosenwald2002use} and contains
240 patients with diffuse large B-cell lymphoma (DLBCL) 
and 7399 genes. The second
dataset has 4751 gene expression measurements of 78
breast cancer patients \citep{van2002gene}. The third
consists of 86 lung cancer patients measured on 7129 genes
\citep{beer2002gene}, and finally, we
analyze a dataset consisting of 116 patients with acute myeloid leukemia (AML,
\citealp{bullinger2004gene}) and 6283 genes.

Since the survival times for some patients are censored and
right-skewed, we use $\log(\textrm{survival time} + 1)$ as the
response. A Cox model would be more
appropriate, but this transformation is enough to illustrate our
methodology. In order to assess our method using limited data, we randomly
select half of the data as the training set and let the rest be in the
testing set, then estimate each model using the training half and
predict the held out data. We repeat this procedure for 10 random splits and
report the average error. We use 10-fold
cross-validation on the training set to choose all tuning parameters
($t_*$, $b_*$, $d$, and $\lambda$ where appropriate), 
mimicking the procedure of a real data analysis. 

We apply 7 methods on each dataset: 1) PCR; 2) lasso; 3) ridge regression;
4) SPC; 5) SPC+lasso; 6) AIMER($b=0$); and 7) AIMER. We use the {\tt
  R} packages {\tt pls}~\citep{MevikWehrens2007}
to perform PCR and \texttt{glmnet}~\citep{FriedmanHastie2010} to perform lasso
and ridge. For PCR, SPC, SPC+lasso, AIMER($b=0$), and AIMER, we allow
the number of components $d$ to be chosen between 1 and 5. 

Our results are shown in \autoref{tab:1}. For each dataset, we show
the MSE on the testing set, the number of selected genes, and the
number of principal components used (if relevant), averaged
across the 10 random training-testing splits. We do not show results
for PCR because it is uniformly awful.
The results in \autoref{tab:1} are largely consistent with the conclusions we
derive from simulations. AIMER and SPC+lasso tend to select a similar number of
genes, though AIMER has better prediction error on 3 of the
4 datasets. Interestingly, the genes selected by SPC+lasso, lasso, and AIMER
rarely overlap, suggesting that to identify genes for further study,
one should try all three methods. The online Supplement lists the
genes identified by AIMER  for each dataset. In the case of DLBCL, we
also list any previous research relating the selected genes to lymphoma.

The Lung
Cancer data is rather odd in that AIMER$(b=0)$ has better performance
than AIMER. This anomaly is likely because, in contrast with the other datasets, the lung cancer
expression measurements have not been scaled relative to a control group. We
tried two transformations using only the treatment group to
approximate such a scaling, but, while the performance of our method becomes comparable
to SPC following transformations, it remains
slightly worse. 
Without a control group, it is difficult to explain
this outcome with any certainty. A comparison of these
alternative transformations with our results in~\autoref{tab:1} is
contained in the online Supplement.

As seen in the table, ridge regression is sometimes the best of all the
methods. Previous experience suggests that ridge regression is
dominant if the genes are highly correlated or when there is not a
particularly predictive set of genes.
However, the fact that ridge does not screen out unimportant genes 
is a barrier to its applications in genomics. 
On the other hand, AIMER approaches or exceeds the small prediction error of
ridge regression while also selecting a small number of predictive genes, 
making it a better candidate for solving these types of problems.

\section{Discussion}
\label{sec:discussion}

High-dimensional regression methods help in predicting future survival time and 
identifying possibly predictive genes for diseases.
However, the large number of genes, the limited access to patients, 
and the complex covariance structure between genes make the problem
both computationally and statistically difficult.
In both simulations and analysis of actual gene expression datasets,
AIMER has comparable or slightly improved prediction accuracy relative
to existing methods and finds small numbers of
actually predictive genes, all while having a similar computational
burden.
On the other hand, there are some issues which warrant further exploration.

A major benefit of SPC is that it comes with theoretical guarantees
under certain assumptions. While our methodology is intended to work when
these assumptions don't hold, we do not yet have comparable
guarantees. However, the simulated experiments in this paper have
suggested how we might derive such results in a more
general setting.

For the real data examples in this paper, we applied a simple
monotonic transformation to the response variable, however, extending
our methods to Cox models, which are more appropriate, and other
generalized linear models for predicting discrete traits is highly
desirable. It may also be useful to examine other eigenstructure
techniques such as Locally Linear Embeddings or Laplacian Eigenmaps to
produce non-linear predictors. Finally, using other matrix
approximation techniques may yield improved performance or be more
amenable to theoretical analysis.

\section*{Funding}

This work is supported by the National Science Foundation [grant
number DMS--14-07439 to D.J.M.].

\vspace{8pt}

\noindent{\em Conflict of interest:} none to declare.

\appendix

\section{Genes identified for the DLBCL data}
\label{genes-DLBCL}

In this supplement, we perform AIMER on all four datasets
(DLBCL~\citep{rosenwald2002use}, breast cancer~\citep{bullinger2004gene}, lung
cancer~\citep{beer2002gene}, AML~\citep{van2002gene}) discussed in the
manuscript. 
Rather than using training sets containing 50\% of the data as in
the main paper, we use all the observations here. We allow our method
to select up to as many features as there are observations.

\subsection{DLBCL}
\label{sec:dlbcl}

For the DLBCL data, we not only list the selected genes, but also
attempt to find any discussion of those genes in existing literature. Our final estimated model uses 49 gene features, which
correspond to 26 genes. To examine the relevance of each selected gene for 
DLBCL, we adopt two approaches. The first endeavors to find literature examining the 
biological connection of the identified gene to any type of lymphoma.
The second lists any reference in the (rather lengthy) methodological literature in
statistics, computer science, and bioinformatics that uses statistical or machine 
learning methods to examine the DLBCL dataset.  

We display our findings for all 26 genes in \autoref{tab:1}. To summarize, 16 out of the 
26 genes have been related to lymphoma in the biological literature, and 19 of them 
have already been identified via statistical techniques developed for the DLBCL dataset. 
While many of the 26 genes have been previously connected to lymphoma in general 
and DLBCL in particular, AIMER does identify 4 genes with symbols ALDH2, CELF2, 
COL16A1, and DHRS9 that have not been previously identified in the biological or 
methodological literature. We note that, while we have made every effort to locate each 
gene, given the large and evolving literature on this topic, those we have been unable 
to locate may have none-the-less been previously studied.

\begin{table}
\centering
\resizebox{\textwidth}{!}{
\begin{tabular}{@{}r l c c c c l@{}}
  \toprule
 & Symbol &  In biology   & Source(s) & In methodology & Source(s) & Name of gene \\ 
  \midrule
  1 & ALDH2 &  $\times$ & & $\times$ & & aldehyde dehydrogenase 2 family (mitochondrial)\\ 
  2 & BCL2 &  \checkmark &  \cite{kramer1996clinical,blenk2007germinal} & \checkmark &  \cite{lossos2004prediction,blenk2007germinal}  & BCL2, apoptosis regulator\\ 
  3 & CCND2 & \checkmark  & \cite{blenk2007germinal}  &  \checkmark & \cite{miyazaki2008gene,lossos2004prediction,blenk2007germinal} & cyclin D2\\ 
  4 & CELF2 &  $\times$ & & $\times$ & & CUGBP Elav-like family  member 2\\   
  5 & COL3A1 &  \checkmark &   \cite{rosenwald2002use,blenk2007germinal}  & \checkmark & \cite{blenk2007germinal}    & collagen type III alpha 1 chain\\ 
  6 & COL16A1 & $\times$ & & $\times$ & & collagen type XVI alpha 1 chain\\ 
  7 & CR2 &  $\times$ & & \checkmark & \cite{ma2007additive,miyazaki2008gene} & complement C3d receptor 2\\ 
  8 & CYP27A1 &  $\times$ & & \checkmark & \cite{zhao2010additive} & cytochrome P450 family 27 subfamily A member 1\\  
  9 & DHRS9 & $\times$ & & $\times$ & & dehydrogenase/reductase 9\\  
  10 & EPHB1 & \checkmark & \cite{asmar2013genome}  & \checkmark & \cite{zhao2010additive} & EPH receptor B1\\  
  11 & ESTs & \checkmark &  \cite{rosenwald2002use} & \checkmark & \cite{liu2010kernel}& ESTs\\ 
  12 & FN1 & \checkmark & \cite{rosenwald2002use,blenk2007germinal} & \checkmark &  \cite{lossos2004prediction,blenk2007germinal}   & fibronectin 1\\ 
  13 & FUT8 & $\times$ &  & \checkmark & \cite{li2015novel} & fucosyltransferase 8\\ 
  14 & IGHM & $\times$ & & \checkmark & \cite{blenk2007germinal,miyazaki2008gene,zhao2010additive} & immunoglobulin heavy constant mu\\ 
  15 & IGKC & $\times$ & & \checkmark & \cite{miyazaki2008gene,zhao2010additive} & immunoglobulin kappa constant\\ 
  16 & IRF4 &  \checkmark & \cite{alizadeh2000distinct,radivojac2008integrated} & \checkmark & \cite{blenk2007germinal,li2015novel} & interferon regulatory factor 4\\ 
  17 & KIAA0233 &  \checkmark & \cite{rosenwald2002use,blenk2007germinal} & \checkmark & \cite{blenk2007germinal}  & KIAA0233 gene product\\ 
  18 & LMO2 &  \checkmark & \cite{natkunam2007oncoprotein,alizadeh2000distinct} & \checkmark & \cite{blenk2007germinal,liu2010kernel,lossos2004prediction} & LIM domain only 2\\ 
  19 & MAPK10 &  \checkmark  & \cite{ying2010frequent} & \checkmark & \cite{blenk2007germinal,liu2010kernel,zhao2010additive} & mitogen-activated protein kinase 10\\ 
  20 & MME & $\times$ &  & \checkmark & \cite{blenk2007germinal}  & membrane metalloendopeptidase\\ 
  21 & MMP2 &  \checkmark & \cite{gouda2014association} & \checkmark & \cite{ma2007additive} & matrix metallopeptidase 2\\ 
  22 & MMP7 &  \checkmark & \cite{matsumoto2008significant} & $\times$ & & matrix metallopeptidase 7 \\ 
  23 & MMP9 &   \checkmark & \cite{sakata2004expression,alizadeh2000distinct}  & \checkmark & \cite{liu2010kernel} & matrix metallopeptidase 9\\ 
  24 & MYB & \checkmark  &   \cite{dai201601910} & \checkmark & \cite{blenk2007germinal} & MYB proto-oncogene, transcription factor\\ 
  25 & SPARC & \checkmark & \cite{meyer2011stromal,brandt2013combined} & $\times$ & & secreted protein acidic and cysteine rich\\    
  26 & VPREB3 & \checkmark & \cite{rodig2010pre} & $\times$ & & V-set pre-B cell surrogate light chain 3\\ 
   \bottomrule
\end{tabular}
}
\caption{\label{tab:1} DLBCL Predictive Genes. AIMER selected 26 genes. We note that while we have made every effort to locate all 26 genes in the literature, a $\times$ should be taken to indicate that we were unable to locate a reference for that gene rather than the stronger conclusion that no one has yet investigated it.}
\end{table}

\subsection{Genes identified for breast cancer, lung cancer and AML data}

As before, we allow 
the maximum number of selected genes be the same as the total number of patients. 
AIMER identifies 78 genes with breast cancer data, 12 genes for lung cancer, 
and 50 genes for the AML dataset. We list the top 20 selected genes for breast cancer in
\autoref{tab:3}, all 12 selected genes for lung cancer in \autoref{tab:4}, and the top 20 selected genes
for AML in \autoref{tab:5}.

\begin{table}
  \centering
\begin{tabular}[t]{@{} ll@{}}
\toprule
  &  Gene  \\
\midrule
1 & Contig47405$\_$RC \\
2 & NM$\_$002964\\
3 & NM$\_$002965\\
4 & NM$\_$005980\\
5 & Contig43983$\_$RC\\
6 & NM$\_$017422\\
7 & NM$\_$002963\\
8 & NM$\_$020974\\
9 & Contig50360$\_$RC\\
10 & Contig55725$\_$RC\\
11 & NM$\_$018265\\
12 & NM$\_$006115\\
13 & AK001423\\
14 & NM$\_$004525\\
15 & Contig38438$\_$RC\\
16 & AL050227\\
17 & NM$\_$014479\\
18 & NM$\_$002421\\
19 & NM$\_$000266\\
20 & NM$\_$006419\\
\bottomrule
\end{tabular}
\caption{\label{tab:3} Top 20 selected genes for breast cancer dataset
  by AIMER. }
\end{table}

\begin{table}
  \centering
\begin{tabular}[t]{@{}ll@{}}
\toprule
  &  Gene  \\
\midrule
1 & D49824$\_$s$\_$at \\
2 & X57809$\_$s$\_$at\\
3 & M17886$\_$at\\
4 & S71043$\_$rna1$\_$s$\_$at\\
5 & M87789$\_$s$\_$at\\
6 & V00594$\_$s$\_$at\\
7 & X98482$\_$r$\_$at\\
8 & M34516$\_$at\\
9 & hum$\_$alu$\_$at\\
10 & HG2873-HT3017$\_$at\\
11 & HG3364-HT3541$\_$at\\
12 & HG3549-HT3751$\_$at\\
\bottomrule
\end{tabular}
\caption{\label{tab:4} 12 selected genes for lung cancer dataset by AIMER. }
\end{table}

\begin{table}
  \centering
\begin{tabular}{@{} l p{5in}@{}}
\toprule
  &  Gene  \\
\midrule
1 & 112298 MSLN mesothelin \\
2 & 111553 GAGED2 G antigen, family D, 2\\
3 & 117339 APOC2 apolipoprotein C-II\\
4 & 330384 SERPINF1 serine (or cysteine) proteinase inhibitor, clade F (alpha-2 antiplasmin, pigment epithelium derived factor), member 1 \\
5 & 330504 TRG@ T cell receptor gamma locus\\
6 & 220502 TCF4 transcription factor 4 \\
7 & 101316 HLA-DRB3 major histocompatibility complex, class II, DR beta 3\\
8 & 109247 TRG@ T cell receptor gamma locus\\
9 & 98472 KIAA0476 KIAA0476 gene product \\
10 & 331153 HLA-DRB3 major histocompatibility complex, class II, DR beta 3\\
11 & 313178 KIAA1165 likely ortholog of mouse Nedd4 WW domain-binding protein 5A\\
12 & 330849 HLA-DRB3 major histocompatibility complex, class II, DR beta 3\\
13 & 107072 NCF4 **neutrophil cytosolic factor 4, 40kDa\\
14 & 114151 HLA-DRB3 major histocompatibility complex, class II, DR beta 3 \\
15 & 246144     ESTs  Highly  similar to CAMP-DEPENDENT PROTEIN KINASE INHIB\\
16 & 103236 TRG@ T cell receptor gamma locus \\
17 & 118267 SDPR serum deprivation response (phosphatidylserine binding protein)\\
18 & 114582 HLA-DPB1 major histocompatibility complex, class II, DP beta 1 \\
19 & 309986 THY1 Thy-1 cell surface antigen \\
20 & 119834 LPHH1 latrophilin 1\\
\bottomrule
\end{tabular}
\caption{\label{tab:5} Top 20 selected genes for AML dataset by AIMER. }
\end{table}

\section{Alternative analysis for lung cancer data}
\label{sec:altern-analys-lung}

Compared with the other three datasets, the public lung cancer data
comes presents gene expression measurements for
only patients who have been diagnosed with lung cancer. The other
three datasets instead give the logarithm of the ratio between  
diseased sample expression measurements and a reference control
group. To try to make the lung cancer dataset comparable to 
the others, we perform two separate transformations on the data. The
first transformation is to take the base-2 logarithm of all the
expression measurements.
Because some measurements are negative, before taking the logarithm, we first add the negative of
the minimum value plus one
to each feature vector, making all measurements at least 1. This
transformation mimics the standard process. The second transformation
orthonormalizes the gene expression matrix. 

We use the same training and testing procedure as in the main paper on the 
original dataset and the two transformed datasets. \autoref{tab:2} shows the corresponding 
prediction MSE, the number of selected
genes, and the number of components used (when necessary) averaged over 10 training-testing
splits. It turns out that both the $\log_2$ transformation and
normalization improves AIMER relative to the other methods. The number
of components used in AIMER also increases.  
The number of selected genes for AIMER
on the $\log_2$ transformed dataset is the same as with the original
dataset, but AIMER selects more genes on the normalized 
dataset. However, even after these two transformations, AIMER is still
not quite as accurate as SPC.
We posit that using the conventional transformation 
with a control group may enhance the results for AIMER.

\begin{table*}
\centering
\begin{tabular}{@{}l r r r c r r r c r r r @{}}
\toprule
& \multicolumn{3}{c}{original dataset}  & \phantom{a}
  &\multicolumn{3}{c}{$\log_2$ transformation} & \phantom{a}   & \multicolumn{3}{c}{normalization} \\
\cline{2-4} \cline{6-8} \cline{10-12} 
Methods  &  MSE  &  \# genes   &  $d$   &&  MSE  &  \# genes   &  $d$  &&  MSE  &  \# genes   &  $d$  \\
\midrule
lasso                      & 0.8159   & 22   &   && 0.8722 & 16 &  & &0.7921 & 20 & \\
ridge                       & 0.7713   & 7129   &   && 0.7594 & 7129 & && 0.7687 & 7129 & \\
\midrule
SPC                        & 0.8344   & 19   & 3   && 0.8268 & 32 &5 && 0.7799 & 22 & 3\\
SPC+lasso              & 0.8436   & 9   & 4   && 0.8376 & 25 &4 && 0.7864 & 19 & 3 \\
AIMER($b=0$)        & 0.9444   & 7129   & 1   && 0.9570 & 7129 &1 && 4.5202 & 7129 & 3\\
AIMER                    & 1.0203   & 13   & 1   && 0.8901 & 13 &2 && 0.8244 & 42 & 4 \\
\bottomrule
\end{tabular}
\caption{\label{tab:2} The MSE on the test set, the number of selected genes, and the
number of principal components used ($d$ if relevant), each averaged
across the 10 random training-testing splits on the three datasets respectively. }
\end{table*}

\bibliographystyle{mybibsty}
\bibliography{SPCA,aimer-supp}

\end{document}